\documentclass[aps,pra,amsmath,twocolumn,amssymb,floatfixng,showpacs,
superscriptaddress,footinbib]{revtex4}
\pdfoutput=1
\usepackage[dvips]{graphics}
\usepackage{bm}
\usepackage{float}
\usepackage{epsfig}
\usepackage{enumerate}
\usepackage{amsmath}
\usepackage{color}
\usepackage{graphicx}
\usepackage[colorlinks=true,linktoc=page,linkcolor=red,citecolor=blue]{hyperref}

\newcommand{\etal}{{\it et al.~}}
\newcommand{\ie}{{\it i.e.,}}

\newcommand\bea{\begin{eqnarray}}
\newcommand\eea{\end{eqnarray}}
\newcommand\beq{\begin{equation}}  
\newcommand\eeq{\end{equation}}

\begin{document}
	
	\title{Emergence of exceptional points and their spectroscopic signature in Dirac semimetal-dirty Superconductor heterojunction} 
	\author{Sayan Jana}
	\email{sayan@iopb.res.in}
		\affiliation{Institute of Physics, Sachivalaya Marg, Bhubaneswar-751005, India}
		\affiliation{Homi Bhabha National Institute, Training School Complex, Anushakti Nagar, Mumbai 400094, India}	
		\author{Debashree Chowdhury}
		\email{debashreephys@gmail.com}
	\affiliation{Centre of Nanotechnology, Indian Institute of Technology Roorkee, Roorkee, Uttarakhand-247667}
	\author{Arijit Saha}
	\email{arijit@iopb.res.in}
	\affiliation{Institute of Physics, Sachivalaya Marg, Bhubaneswar-751005, India}
	\affiliation{Homi Bhabha National Institute, Training School Complex, Anushakti Nagar, Mumbai 400094, India}		
	\date{\today}
	\begin{abstract}
	We theoretically investigate the emergence of non-hermitian physics at the heterojunction of a type-II Dirac semi-metal (DSM) and a dirty superconductor (DSC). The non-hermiticity is introduced in the DSM through the self-energy term incorporated via the dirtiness of the superconducting material. This causes the spectra of the effective Hamiltonian to become complex, which gives rise to the appearance of the exceptional points (EPs). This complex self energy, apart from having a frequency dependence, also acquires spatial dependence as well, which is unique and can provide interesting effects related to non-hermitian physics in spectral function analysis. At an appropriate distance from the normal metal-superconductor junction of the DSC, non-hermitian degeneracies appear and a single Dirac point splits into two EPs. In the spectral function analysis, apart from the EPs, a Fermi-arc like structure also emerges, which connects the two degeneracies (EPs). The results discussed here are distinctive and possibly can be realized in spectroscopy measurements. 
	\end{abstract}
	\maketitle
	\section{Introduction} During the last decade, the notion of topology has attracted a lot of attention in the field of modern condensed matter physics~\cite{Kane,Hasan,Vish}. 
The recent focus in this direction are the topological semimetals (TSMs), the commonest example of which is the Dirac semimetal (DSM) and Weyl semimetal (WSM)~\cite{Armitage}. In WSM, the two minimal non-degenerate bands touch at topologically protected points, coined as Weyl points. On the other hand, 
in Dirac semimetals (DSMs), bands meet at a single Dirac point. In contrast to WSM, in DSM the Dirac point is protected by both time-reversal (TR)~\cite{Burkov} and inversion (I)~\cite{Halasz} symmetries.
By breaking either (TR/I) or both one can transform DSM into WSM~\cite{Armitage}. It is important to note that in TSMs, any hermitian perturbation can only shift the position of the special points 
(Dirac/Weyl), but is unable to crate or destroy them.
\begin{figure}
	\includegraphics[width=.9 \linewidth]{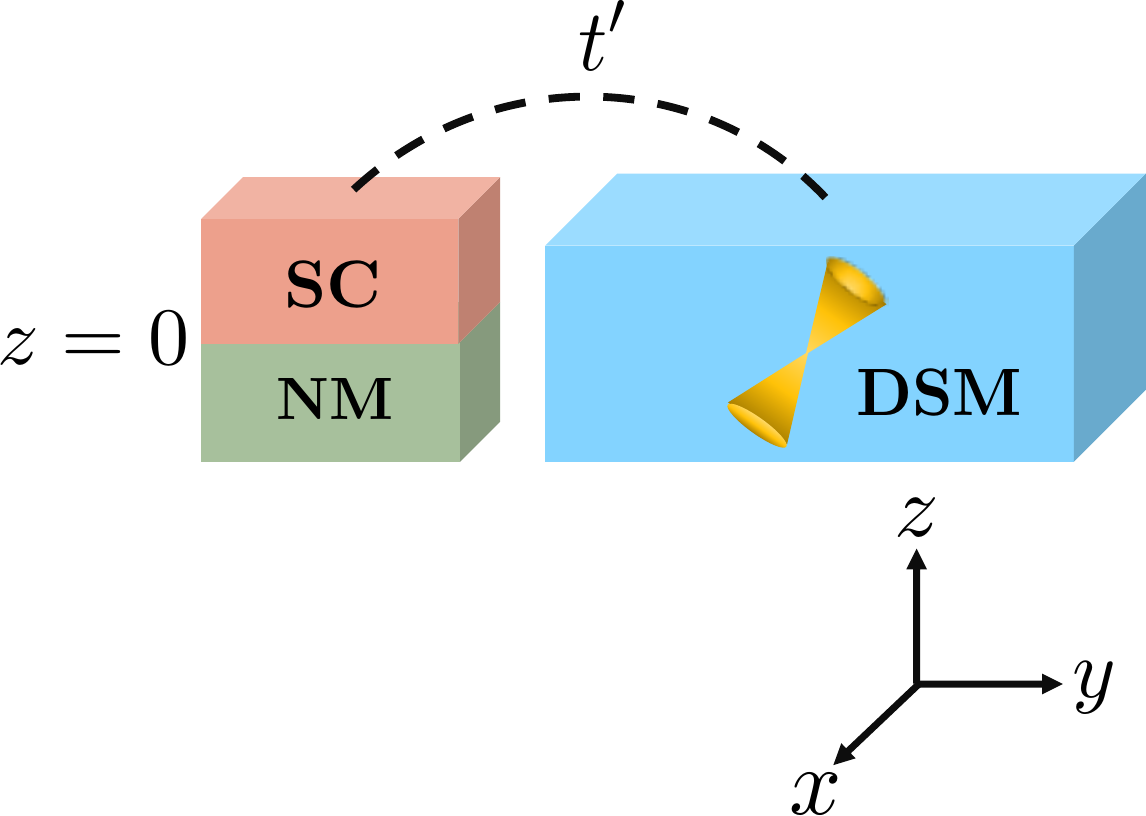}
	\caption{(Color online) A schematic diagram of our setup is demonstrated. We consider a junction of type-II DSM (light blue, light grey)-DSC (green, orange, light grey) where $t^{\prime}$ is the tunneling amplitude between them. DSC is a diffusive junction at $z=0$ between a normal BCS superconductor (orange, grey) and a normal metal (light green, light grey). In type-II DSM, the Dirac cone is tilted as schematically shown here. 
	}
	\label{fig1}
\end{figure}
Besides the interesting topological aspects discussed above in case of hermitian systems, the new challenge in this field is to explore the appearance of topological phases in presence of non-hermitian (NH) perturbations~\cite{Yao,Lee,Xu,Shen,Shen1,Papaj}. In photonic systems, perturbations like material gain/loss, radiative out-coupling induce non-hermiticity~\cite{Ganainy}. This opens up a new window towards the analysis of several unique features related to loss-induced transmission~\cite{18,19}, single mode operation~\cite{11,12,13,14,15,16,17} and reverse pump dependence in lasers~\cite{20,21}. 
Apart from these studies in photonics, the community is also interested in analyzing the role of such 
loss/gain in topological systems~\cite{Yao,Lee,Xu,Shen,Shen1,Papaj,Bergholtz1,Ashida,Denner} including Majorana fermions~\cite{Jose,Avila}. From the quantum mechanics point of view, non-hermiticity is not an emerging new concept. In literature, the ${\mathcal {PT}}$ symmetric NH Hamiltonians were analyzed long 
before~\cite{Bender}, where the energy spectra can still be real without any hermiticity constraint. In recent times, the attention has been shifted towards the effective description of the Hamiltonian~\cite{Torres} for systems having finite lifetime introduced by electron-electron (e-e)~\cite{ee}, electron-phonon (e-ph) \cite{Kozii} interactions or by incorporating disorders~\cite{Zyuzin,Zyuzin1}. Finite lifetime of the system renormalizes the low-energy dispersion of materials in an effective way by introducing the self energy term in the single particle retarded Green's function as~\cite{Bergholtz} $G^{R}(\omega)=(\omega-H(k,\omega))^{-1}$ with $H=H_{0}(k)+\Sigma(\omega)$.
	
	Here, $H_{0}(k)$ is the single particle Hamiltonian of the system under consideration and $\Sigma(k,\omega)$ is the self-energy of the electron appears due to interactions (e-e/e-ph) or electron-impurity scattering~\cite{Zyuzin, Zyuzin1}. The single particle Hamiltonian $H_{0}(k)$ is Hermitian in this case. On a contrary, the self energy term $\Sigma(k,\omega)$ turns out to be imaginary in presence of finite lifetime contributions. Thus, it translates the entire problem in the effective quasi-particle picture to be NH and the band structure becomes complex. The remarkable feature of this NH system is that its eigen-values exhibit special degenaracies at certain momentum values within the Brillouin zone (BZ), which are coined as the exceptional points (EPs) in the literature~\cite{Ashida,Bergholtz1}. At these points the Hamiltonian of the system becomes defective and the wave-functions merge~\cite{Bergholtz1}. The EPs are the degenerate points where the vorticity (a topological invariant)~\cite{Torres} becomes quantized at $\pm 1/2$.

In this article, we are proposing a new route to introduce non-hermiticity in a type-II DSM~\cite{Armitage} via the self-energy term incorporated by a diffusive (dirty) superconductor (DSC). This has not been explored so far, in the context of NH phenomena, to the best of our knowledge. The DSC introduces imaginary self-energy term to the DSM and consequently switch the whole problem in the effective single particle picture from hermitian to NH. 
We obtain the condition for the appearance of EPs within our set-up and also manifest the spectral function signature of the EPs based on continuum as well as microscopic lattice model.  

The remainder of our paper is organized as follows. In Sec.~\ref{secII} we discuss our model Hamiltonian and the method used in this paper. Sec.~\ref{secIII} includes the main results of the paper. It consists of three subsections. In them, the detailed analysis of the apearence of the EPs in the continum model, spectral function and the lattice model details are incorporated respectively. Finally we conclude in Sec.~\ref{secIV} with a summary of our results.

\section{Model and Method \label{secII}} 
We begin with a heterojunction that consists of a type-II (tilted) DSM in close proximity to a DSC, which is a 
superconductor-normal (SN) bilayer, where the electrons from the normal metal diffuse into the superconductor~\cite{Danon}. The schematic diagram of our model is depicted in Fig.~\ref{fig1}. 

The low energy effective Hamiltonian of DSM having linear dispersion around a single gap-less Dirac point is given by~\cite{Armitage,Zyuzin1}
\begin{eqnarray}\label{1} 
H_{DSM}={\mathfrak T} v k_{x}\sigma_{0}+\hbar v (\bm\sigma\cdot \bm k-\mu I_{2*2}),
\end{eqnarray}
where $\bm \sigma$ and  $\mu$ are the Pauli matrices in the spin space and the chemical potential respectively. ${\mathfrak T}$ is the tilt parameter signifies the inclination of the Dirac cone along 
$k_{x}$ direction. In Eq.(\ref{1}), $v$ is the Fermi velocity and $\sigma_{0}$ denotes the $2\times 2$ unit matrix. Inclusion of the tilt parameter in the system breaks the particle-hole symmetry. 
The tilt is an essential parameter for capturing the NH phenomena~\cite{Zyuzin1}. 

The features of the diffusive superconductor is incorporated in DSM through the self energy term of the superconductor $\Sigma_{SC}^{R}(\omega,z,\gamma)$. Note that, the parameter 
	\begin{equation}\label{10}
	\gamma=(\sigma_n\xi_s/\sigma_s\xi_n)\; ,
	\end{equation}
	is important as it signifies the mismatch in the coherence lengths and conductivities of the two materials of the SN bilayer. Here, $z$ corresponds to the distance inside the superconductor. Also, 
$\sigma_{s(n)}$ and  $\xi_s(\xi_n)$ are the conductivity and
	coherence length of the superconductor (normal metal) respectively.

	By integrating out the superconducting degrees of freedom one obtains~\cite{Danon},
\begin{eqnarray}
\Sigma_{SC}^{R}(\omega,z,\gamma)=|t^{'}|^{2} G^{R}_{SC}(\omega,z,\gamma),
\end{eqnarray}
where $ G^{R}_{SC}(\omega,z,\gamma)$ is the retarded Greens function of the superconductor and $t^{'}$ is the tunneling parameter. Now to calculate the electron, hole and anomalous parts of the Greens function, we follow Refs. \cite{Danon,Belzig}, where
one usually employs the well known semi-classical Usadel equation to solve the dirty SN junction problem with 
$\Delta(r)=\Delta$. At the interface ($z=0$), using the angular parametrization for the electronic and anomalous part of the Greens functions as $G_{ee}=\cos(\theta),F_{eh}=\sin(\theta),$ the solution for the Usadel equation (in the thick limit) can be obtained as~\cite{Danon,Belzig} 
\begin{widetext}
	\begin{align}\label{8}
	\theta(\omega,z,\gamma)=\left\{
	\begin{array}{l}
	4\text{arctan}[\;\tan(\theta_0/4)
	\exp(\sqrt{-2 i \omega/D_n}\;z)\;] \quad \hfill z<0 
	\\
	\theta_s+4\;\text{arctan}[\;\tan((\theta_0-\theta_s)/4)
	\exp(-\sqrt{2\sqrt{-\omega^2+\Delta^2}/D_s}\;z)\;]
	\quad \hfill ~~z>0.
	\end{array}\right\},
	\end{align}
\end{widetext}
where $D_{s(n)}$ is the diffusion constant of the superconductor (normal metal).
Note that, we introduce the DSM not on the top of the bilayer structure, rather the last layer of the DSM in the horizontal plane ($xy$) is coupled to a particular distance of the bilayer structure by the tunnel coupling term $t^{\prime}$. In principle, this tunnel coupling term $t^{\prime}(z)$ can also be position dependent. Although, in our case  we neglect this position dependence of the coupling parameter and consider $t^{\prime}(z)$=$t^{\prime}$ for simplicity. In Eq.(\ref{8}), 
\begin{eqnarray}\label{9}
  \theta_s & = & \text{arctan}\Big(\frac{\Delta}{-i\omega}\Big),\\
  \sin\frac{\theta_0-\theta_s}{2} & =&
  \gamma\frac{(-i\omega)^{1/2}}
  {(-\omega^2+\Delta^2)^{1/4}}
  \sin\frac{\theta_0}{2}\;.
\end{eqnarray}

\begin{figure}
	\includegraphics[width=1.05 \linewidth]{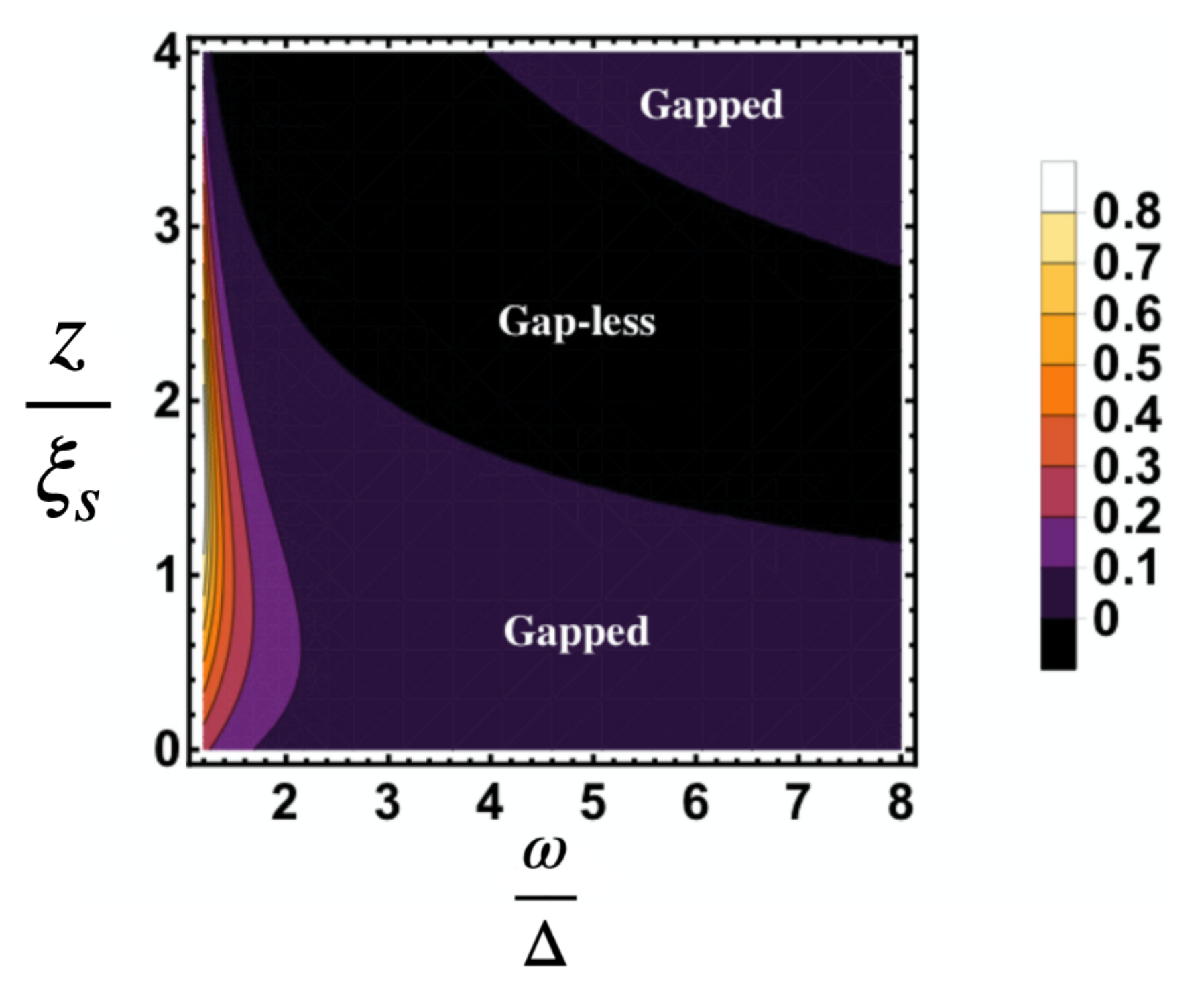}
	\caption{(Color online)   
		A phase diagram is demonstrated in the plane of energy ($\omega$) and distance ($z$) within the DSC. Here the background color manifests the magnitude of the energy gap (at $k_{x}=k_{y}=0$) controlled by the imaginary part of the self-energy of the DSC. The violet (dark grey) region corresponds to the trivial gapped phase and the black region (dark grey) belongs to the gap-less phase of EPs. This phase diagram carries information about the critical thickness of the normal superconductor from the junction (SN) at $z=0$ (see text for discussion). Here we have choosen $\gamma=1.$}
	\label{fig2}
\end{figure}
Thus the self-energy of the DSC reads~\cite{Danon}
\begin{align}\label{13SE}
\Sigma_{SC}^{R}(\omega,z,\gamma)=\zeta (\sin\theta(\omega,z,\gamma) ~\tau_{y}-i\cos\theta(\omega,z,\gamma) )I_{2\times 2},
\end{align} 
where $\zeta\sim\lvert t^{'}\rvert^{2}$ denotes the coupling between the
DSM and DSC at the junction and $\tau$ matrices act on the particle-hole space.
A similar analysis for a thin superconducting layer is reported in Ref.~\cite{Fominov}, where an experimental set up for such SN bi-layer structure is discussed.

It is evident from Eq.(\ref{10}) that $\gamma=0$ signifies the absence of the normal metal and yields the usual bulk $s$-wave superconductor effects with the following form of the 
self-energy~\cite{Danon}
\begin{align}\label{14}
\Sigma_{SC}^{R}(\omega)=\frac{\zeta}{\sqrt{\Delta^2-\omega^2}}(\Delta~\tau_{y}-\omega I )
I_{2\times 2}.
\end{align}
Using Eqs.(\ref{1}) and (\ref{13SE}), the effective quasi-particle Hamiltonian, which we define from the retarded Green's function of the system, reads
\begin{align}\label{15}
H_{\rm eff}=H_{DSM}+\Sigma_{SC}^{R}(\omega,z,\gamma).
\end{align}
This recasts the problem in a more unique and interesting format
from the perspective of NH exceptional physics.

\section{Results \label{secIII}}
\subsection{Appearance of EPs} The DSC induces a position and frequency dependent self-energy 
in the DSM system and as a result the  effective Hamiltonian (Eq.(\ref{15})) becomes NH due to the contribution originating from the complex part of the self energy $\Sigma_{SC}^{R}(\omega,z,\gamma)$.
Therefore, the eigen-values for the full effective Hamiltonian for a particular frequency $\omega$ in the 
($\mu\to 0$) limit is given as
\begin{align}\label{17}
&{\cal E}_{\pm}^{}(\omega,k,z,\gamma)=  {\mathfrak{T}} v k_{x} -i \zeta  \cos (\theta(\omega,z,\gamma) )\nonumber\\&\pm\sqrt{-\frac{\zeta ^{2}}{2} \cos(2\theta(\omega,z,\gamma) )+\frac{\zeta ^{2}}{2}+\hbar^{2} v^{2} \left(k_{x}^{2}+k_{y}^{2}\right)},
\end{align}
\begin{figure*}[!bthp]
		\hspace*{\fill}%
	\includegraphics[width=1.05 \linewidth]{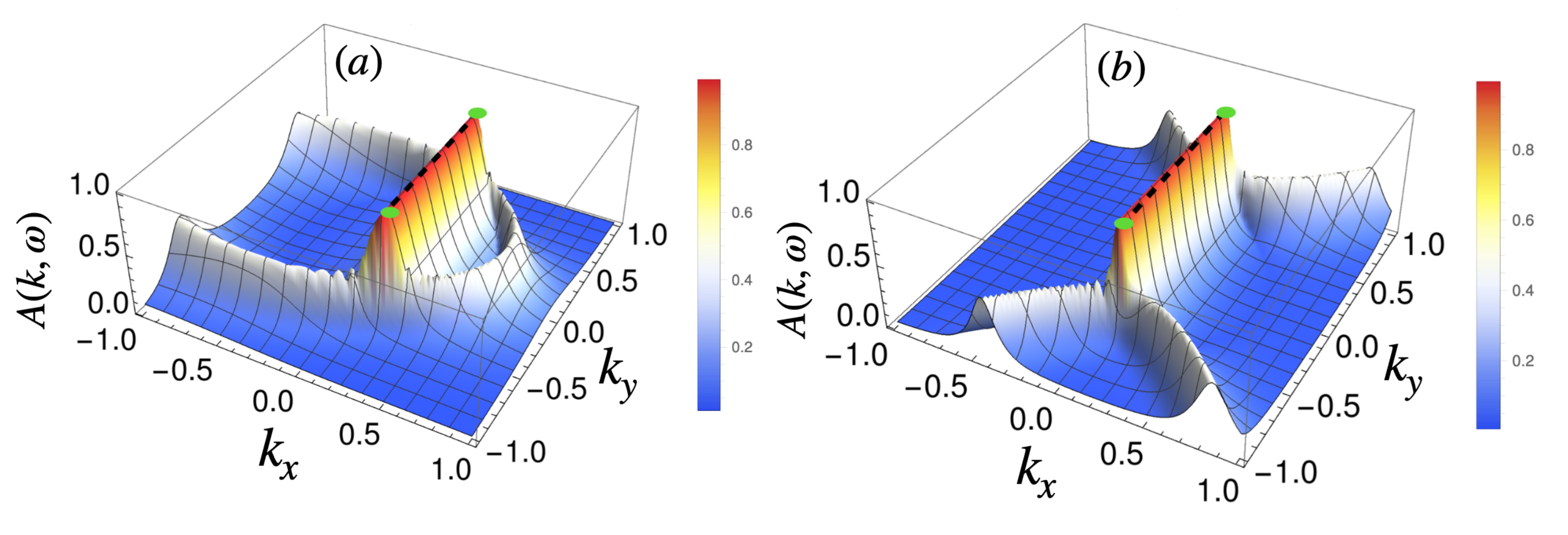}\hfill%
		\hspace*{\fill}%
          \vspace {-0.5 cm}
	\caption{(Color online) The momentum-resolved spectral function $A(k,\omega)$, for the contnuum model (Eq.(\ref{1})), is illustrated for a given frequency $\omega=2\Delta$, in $k_{x}$-$k_{y} $ plane. The emergent Fermi-arc structure discussed in the text is shown by the black dotted lines that connect the two green dots denoting the EPs. For panel (a) we choose the tilt parameter ${\mathfrak T}=0.8~(< 1)$ and in panel (b) the tilt is chosen to be ${\mathfrak T}=1.6~(>1)$. The other parameters are chosen as $t^{\prime}=0.8,\gamma=1, z=2.6\xi_{s}$. One observes that large tilt enhances the length of the Fermi-arc connecting the two EPs.}
\label{fig3}
\end{figure*}
where we consider $k_{z}=0$ plane (here $k_{z}$ is a good quantum number in our
continum low-energy model of the DSM. The position dependence ($z$) of the spectrum arises 
due to the distance within the superconductor \ie~via the superconducting self-energy). The eigen-values in Eq.(\ref{17}) are complex due to the $\theta(\omega,z,\gamma)$ dependence, which itself is complex, as is evident from Eq.(\ref {8}).

Note that, when the frequency dependent eigen-values become degenerate \ie~${\cal E}_{+}^{}(\omega,z,\gamma)={\cal E}_{-}^{}(\omega,z,\gamma)$ one observes exceptional degenaracies (EPs). At these EPs not only eigen-values but also the eigen-vectors of the effective Hamiltonian coalesce. The complex square-root singularity in the eigen-values within Eq.(\ref{17}) is a specific property of the EPs. The exact conditions for generating these EPs can be found following~Ref.~\cite{Yoshida}
\begin{align}\label{EP}
\frac{\zeta^{2}}{2}\Big({\rm Re}[\cos(2\theta(\omega,z,\gamma))]-1\Big)&=\hbar^{2}v^{2}k^{2},\nonumber\\
{\rm Im}[\cos(2\theta(\omega,z,\gamma))]&=0,\nonumber\\
\omega&={\mathfrak{T}} v k_{x}.
\end{align}

In Fig.~\ref{fig2}, we show the variation of the energy gap (${\cal E_{+}} - {\cal E_{-}}$), at $k_{x}=k_{y}=0$ and controlled by ${\rm Im}[\cos(2\theta(\omega,z,\gamma))]$, as a function of $z$ and $\omega$ and $\gamma.$ Fig.~\ref{fig2} shows that for $\gamma=1,$ EPs appear in the quasiparticle spectrum when $\omega> \Delta$ and $z> 1.$  
At these parameter values ${\rm Im}[\cos(2\theta(\omega,z,\gamma))]$ becomes zero at certain region, which contains the EPs. On the other hand, for $\omega < \Delta$ (see Eq.(\ref{13SE})) we always encounter a gap in the spectrum as  ${\rm Im}[\cos(2\theta(\omega,z,\gamma))]\neq 0$ and this causes no special degenaracies. 

\begin{figure}[!thpb]
	\centering
	\includegraphics[width=1.05 \linewidth]{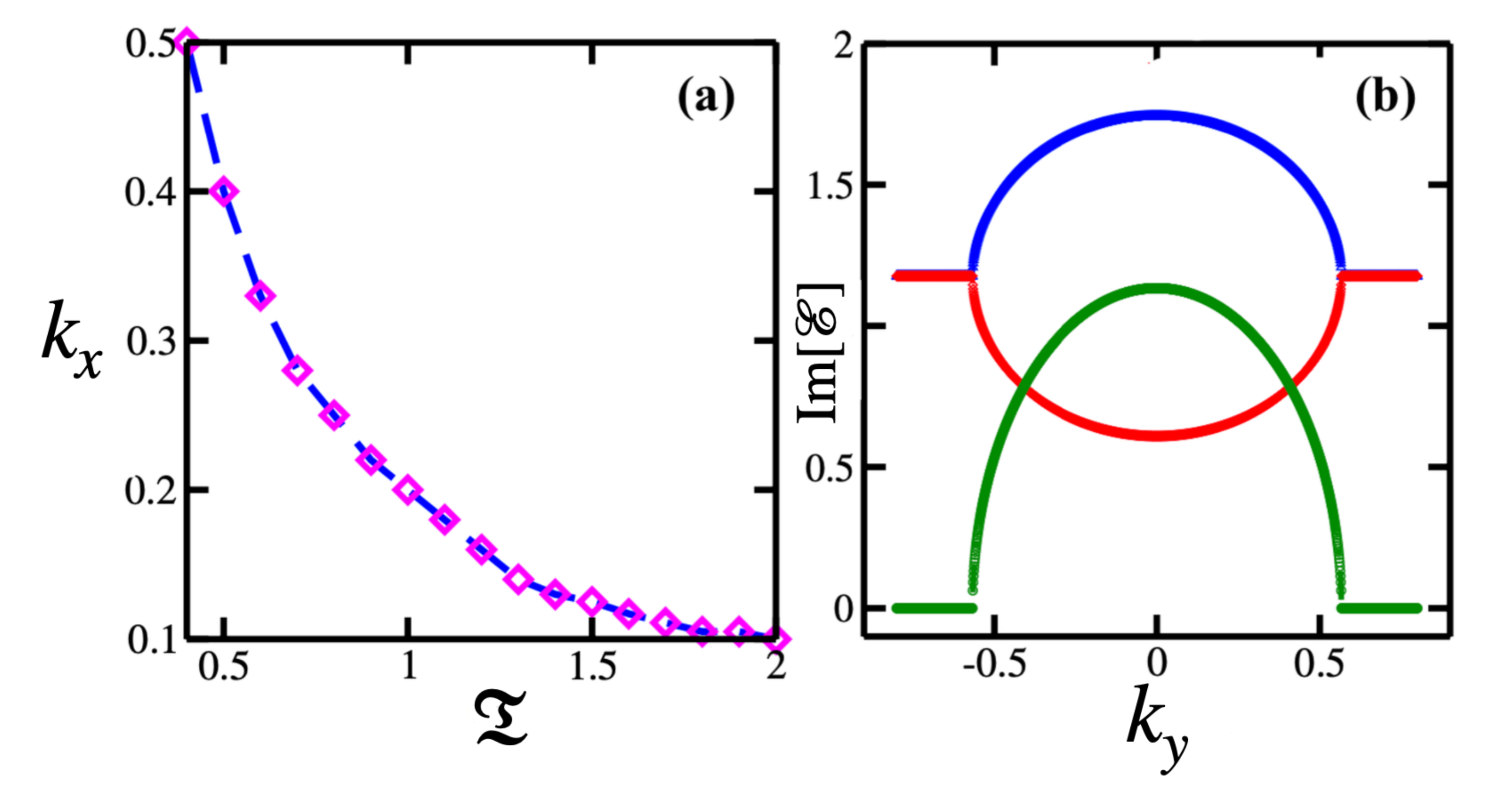}
	\caption{(Color online) (a) The change in the position of EPs along the $k_{x}$ direction is depicted
as a function of tilt ${\mathfrak T}$. Enhancement in the magnitude of the tilt parameter decreases the value of $k_{x}$, for chosen parameter $\omega=2\Delta, \gamma=1.0$. (b) The imaginary parts of complex 
eigen-values (blue and red lines) are shown as function of $k_{y}$ for the effective Hamiltonian 
(Eq.(\ref{15})). We choose the values of the other parameters as $\omega=2\Delta$,~$k_{x}=0.25$, 
${\mathfrak T}=0.8$ and $\gamma=1.0.$ Here the green line exhibits the difference between these two values. This difference becomes zero at the transition points (EPs).}
	\label{fig4}
\end{figure}

\subsection{Spectral Function analysis} The retarded Green's function of the system can be defined as 
\begin{align}\label{18}
G_{sys}^{R}(k, \omega,z,\gamma)=\frac{1}{[\omega-H_{\rm eff} + i\eta]},
\end{align}
where $\eta$ denotes the broadening parameter. With the help of the Green's function one defines the spectral function as~\cite{Yoshida},
 \begin{align}\label{19}
A(k,\omega,z,\gamma)=-\frac{1}{\pi}\sum_{s=\pm 1}\Big(\omega+i\eta-{\cal E}_{s}(\omega,k,z,\gamma)\Big)^{-1}.
\end{align}
	\begin{figure*}[!bthp]
	\hspace*{\fill}%
	\includegraphics[width=1.05 \linewidth]{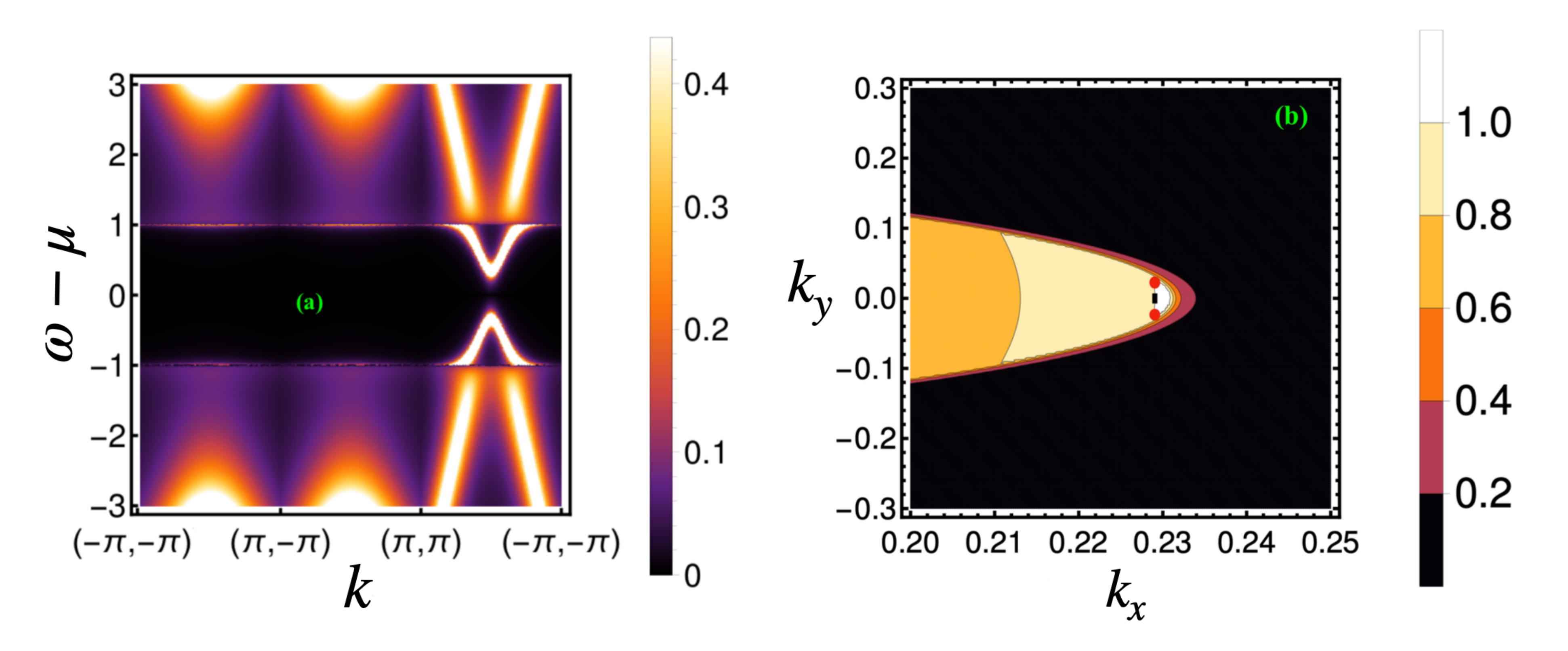}\hfill%
	\hspace*{\fill}%
         \vspace {-0.5 cm}
	\caption{(Color online) (a) Considering the continuum model (Eq.(\ref{1})), the momentum resolved spectral function $A(k, \omega)$ is demonstrated along the path $k$ ($k_{x},k_{y}$)=(-$\pi$,-$\pi$) to ($\pi$,-$\pi$) to ($\pi$,$\pi$) and again back to (-$\pi$,-$\pi$) for tilt $\mathfrak{T}$=0. Discontinuity at 
$\omega=\pm\Delta$ clearly manifests that for $\mathfrak{T}=0$ there is no gap closing in the spectral function. This signifies the absence of EPs in this case. (b) 
$A(k,\omega)$ is depicted in the $k_{x}$-$k_{y}$ plane based on the microscopic lattice model 
(Eq.(\ref{lm1})). Here, we choose $\omega=2\Delta$ and $\mathfrak{T}=0.8$. 
The maximum spectral weight signifies the existence of EPs (red dots) and the arc-structure (black dotted line) in between them. The other parameters of the lattice Hamiltonian are chosen as t=0.5 $\lambda=1$, 
$\lambda_{z}=1$ $\tilde{t}=0.8t$ and $M=5.9t$ (to be in the DSM phase).
	}
      \label{fig5}
	\end{figure*} 
To analyze the resultant-spectral function signature of the EPs in the quasi-particle spectrum 
($\omega=2 \Delta$), we present momentum resolved spectral function $A(k, \omega)$ as a function of $k_{x}$ and 
$k_{y}$ in Fig.~\ref{fig3}(a)-(b). First, we illustrate our results based on the continuum model of the DSM 
(Eq.(\ref{1})).
By following the path of maximum spectral weight, one observes two end points (EPs) that are connected by an arc like structure given by $\rm Re[{\cal E}]=0$. Note that, the weight of the spectral function gradually decreases due to the gapped quasiparticle spectrum as we move away from the arc in momentum space. 
This is shown in Fig.~\ref{fig3}(a). In a different scenario, when the tilt parameter $(\mathfrak{T}>1=1.6)$ is greater than the Fermi velocity, the EPs exist with an enhanced connecting length of the arc along the 
$k_{y}$ direction. The situation is depicted in Fig.~\ref{fig3}(b). To study the robustness of the EPs as a function of the tilt, we depict the position of them along $k_{x}$ direction as a function of tilt in 
Fig~\ref{fig4}(a). As a matter of fact we observe that the $k_{x}$ value decreases with the enhancement in the tilt parameter $\mathfrak{T}$ for $\omega=2 \Delta$. The length of the arc thus must increases in the $k_{y}$ direction as can be seen from the condition in Eq.(\ref{EP}). In Fig. \ref{fig4}(b) we have analyzed the Fermi-arc behavior, which is related to the imaginary parts of the ${\cal E}_{\pm}^{}(\omega, k, z,\gamma)$. 
The imaginary parts of the eigen-energies exhibit differences in the values although the real parts still stay degenerate. For a given distribution (Gaussian or Lorentzian) the imaginary part of the self-energy is related to broadening or inverse of life time. 
Therefore, difference in the imaginary parts of the ${\cal E}_{\pm}$ which is related to $\Sigma_{SC}^{R}(\omega,z,\gamma)$ (Eq.(\ref{13SE})) resulting in different life time for the two eigen-energies and that is the source of the Fermi-arc structure. At the EPs, ${\rm Im}({\cal E}_{\pm})$ becomes equal indicating equal half life. Thus the measurement of life-time can be a possible tool to analyze the Fermi-arc structure connecting the two EPs. Note that, both the DSC induced $\theta(z,\omega,\gamma)$ term (in ${\cal E}_{\pm}$) and the particle hole symmetry breaking tilt term ${\mathfrak T}$ in DSM play equally important role in the formation of EPs. To elaborate the role of $\mathfrak{T}$, we show the feature of $A(k, \omega,\gamma)$ in Fig.~\ref{fig5}(a)
for $\mathfrak{T}$=0. In this case, one can see a clear discontinuity at $\omega=\pm\Delta$, which 
manifests that for $\mathfrak{T}=0,$ there is no gap closing in the spectral function. Hence, no EPs appear here for $\mathfrak{T}=0.$ .

\subsection{Lattice Model}
Finally, to support our continuum results, we analyse a microscopic lattice model based on three-dimensional topological insulator following Ref.~\cite{Lattice}. The 4 band Hamiltonian for the cubic lattice in momentum space can be written as,
\begin{align}\label{lm1}
H_{\rm TI}&=  2\lambda\sigma_{z}(s_{x}\sin k_{y}-s_{y}\sin k_{x})+2\lambda_{z}\sin k_{z}\sigma_{y}\nonumber\\&+[M-2t(\cos k_{x}+\cos k_{y} +\cos k_{z})]\nonumber\\&+\mathfrak{T}v\sin k_{x}\ .
\end{align}
Here, $s_{i}~(\sigma_{i})$ denotes the Pauli matrix in orbital (spin) space in the $i^{\rm th}$ direction. 
$M$ is the Dirac mass parameter, $t$ is the nearest neighbor hopping amplitude and $\lambda, \lambda_{z}>0$ are the spin orbit coupling strength. The Dirac mass should be within the range $(2t<M<6t)$ to realize a DSM phase within this model~\cite{Lattice}.

The tunneling Hamiltonian between the bottom layer of the DSM (at $z_{0}$) and a specific layer of DSC (at z) can be written as~\cite{Lattice1,Lattice2}
\begin{align}\label{lm2}
H_{\rm Tunnel}&=\sum_{\sigma,\bar{\sigma},s} V_{c}\Psi^{\dagger}_{z_{0},s,\sigma}\Phi_{z,\bar{\sigma}}+h.c\ ,
\end{align}
where, $\Psi$ and $\Phi$ are the wavefunctions of the electron in the DSM and DSC regions respectively. Here, $s$ $(\sigma)$ is the Pauli matrix in orbital (spin) space for the DSM and $\bar{\sigma}$ is the spin index for the DSC.

One can write the coupling term $V_{c}$ that connects the last layer of DSM (at $z_{0}$) with a specific layer of superconductor (at z) as
\begin{align}\label{lm_4}
V_{c}=
\begin{bmatrix}
\tilde{t}_{1} & 0 & 0 & 0\\
0 & \tilde{t}_{1} & 0 & 0\\
\tilde{t}_{2} & 0 & 0 & 0\\
0 & \tilde{t}_{2} & 0 & 0\\
0 & 0 & -\tilde{t}_{1} & 0\\
0 & 0 & 0 & -\tilde{t}_{1}\\
0 & 0 & -\tilde{t}_{2} & 0\\
0 & 0 & 0 & -\tilde{t}_{2}\\
\end{bmatrix} \ ,
\end{align}
where we consider same coupling strengths ($\tilde{t}_{1}$=$\tilde{t}_{2}$) for both the orbitals.

The retarded Green's function of the superconducting bi-layer can be written as,
\begin{align}\label{13}
G_{SC}^{R}(\omega,z,\gamma)= (\sin\theta(\omega,z,\gamma) ~\tau_{y}-i\cos\theta(\omega,z,\gamma) )I_{2\times 2}\ .
\end{align}
Thus one can write down the self-energy as
\begin{align}\label{lm4}
\Sigma_{SC}(\omega,z,\gamma)&=V_{c}G^{R}_{SC}{V_{c}}^{\dagger}\ ,
\end{align}
Hence, the Green's function for the system can be written as,
\begin{align}\label{lm3}
G_{sys}^{R}(k, \omega,z,\gamma)=\frac{1}{[\omega+i\eta-H_{DSM}-\Sigma_{SC}(\omega,z,\gamma)]} \ ,
\end{align}
Therefore, numerical computation of Green's function at each site contains 
(8$\times $8) matrix consisting of (spin, orbital and particle-hole) degrees of freedom. Note that,
we haven't consider any microscopic lattice model for our DSC. Rather we employ the discretized version of the self-energy in order to include the effect of DSC in our lattice model 
(Eq.(\ref{lm1})).

For our numerical analysis, we consider thirty layers of DSM along the $z$ direction starting from $z_{0}$ and compute the layer dependent spectral function as 
\begin{align}\label{lm5}
A^{\alpha}(k,\omega,z)=-\frac{1}{\pi}\sum_{\beta}|\Big <\chi_{\beta}|\alpha\Big >|^{2} \Big(\omega+i\eta-{\cal E}_{\beta}(\omega,k,z)\Big)^{-1},
\end{align}
where $\alpha$ is the layer index, which contains the information of spin, orbital and particle-hole in each layer. Here the momentum indices $k$ is for the $(k_{x},k_{y})$. We consider open boundary condition along $z$ and thus $k_{z}$ is no longer a good quantum number and translational symmetry is broken along $z$. The index $\beta$ corresponds to the number of eigen-values and eigen-vectors 
at each layer and $\Big\{|\chi_{\beta}\Big >\Big\}$ and $\Big\{{\cal E}_{\beta}\Big\}$ denotes the corresponding eigen-vectors and eigen-values of the effective Hamiltonian ($H_{DSM}+\Sigma_{SC}(\omega,z,\gamma)$).

In Fig.~\ref{fig5}(b), the behavior of $A(k, \omega)$, based on the lattice model analysis, manifests similar signature of the EPs and the Fermi-arc structure within the bulk as obtained from the continuum model discussed before.

\section{Summary and Conclusions \label{secIV}} 
To summarize, in this article, the role of the self-energy induced by a DSC on a tilted DSM is analyzed. Importantly, the topological nature of the DSM is remarkably modified in presence of the DSC. The 
self-energy term, induced by the DSC, converts the whole problem to be NH in the effective description. The effective Hamiltonian thus obtained is defective and as a consequence is non-diagonalizable. This defectiveness is a key signature of the NH systems. The obtained NH self-energy splits the Dirac point of the DSM into two EPs with vorticity $\pm 1/2$. In the spectral function analysis, one can observe a peak at the EPs with a Fermi-arc like structure connecting them. Note that, the length of the Fermi-arc structure increases as we enhance the tilting term. One important aspect of our results is that here the defective points arise in the regime where $\omega> \Delta$ \ie~in the quasi-particle spectrum. In the reverse direction, with same order of other parameters, the results could not be reachable. 
Moreover, another important aspect of our work is that the self-energy term induced in the DSM via the DSC exhibits a space dependence, which is an unique feature that contributes to the appearance of EPs. 
Both the tilting parameter and the space dependent self-energy are equally necessary ingredients to 
obtain the EPs within our set-up.  

In a recent experiment, similar NH arc structure has been analyzed~\cite{Zhou1}, where a periodic photonic crystal is signed by a lossy radiation. The appearance of double Riemann sheet topology here leads to the distinct feature of complex Fermi-arc structure between the EPs. These experiments lead to the direct experimental validation of NH band theory. Therefore, by considering the experimental progress in the NH systems, we hope that our proposed results may also be experimentally verifiable for the realization of EPs and the Fermi-arc structure in fermionic systems based on our DSC-DSM heterojunction. In particular, the presence of the EPs enhances the local density of states and as a consequence exhibits significant enhancement in spectral weight. This may be
possible to measure in Angle-resolved photoemission spectroscopy (ARPES) experiments.
 The typical value of lattice spacing in DSM ($\rm Cd_{3}As_{2}$)$\sim$10$~\rm \AA$
~\cite{HemianYi}. In our setup, the EPs are separated in momentum space $\delta k \sim 0.01~\rm \AA^{-1}$, the spacial distance corresponding to this momentum spacing would be of the order of $\delta R \sim 10~\rm nm$. Hence, the finite size gap would be $\delta E \sim 90~\rm meV$. Therefore, the diameter of the DSM nanowire needs to be chosen greater than $\delta R$ ($\approx 20-30~\rm nm$) to avoid the finite size quantization and surface effects. The separation of the EPs in momentum space ($\delta k \sim 10^{-2}~\rm \AA^{-1}$) can be within the resolution of the Laser ARPES ($10^{-3}$ - $10^{-4}$ $\rm \AA^{-1}$)~\cite{Iwasawa}. The typical values of the other parameters in our lattice model can be $v_{f}\sim 1.3\times 10^{6}~\rm m/s$, $\mathfrak{T} \sim 0.8\times 10^{6}~\rm m/s $, $t\sim(1-4)~\rm eV, \lambda\sim(0.6-2)~\rm eV, \lambda_{z}\sim(0.6-2)~\rm eV, 
M\sim (5-20)~\rm eV$ etc.

{\bf{\textit{Acknowledgments:}}} We acknowledge the use of the SAMKHYA: High Performance Computing Facility provided by IOP, Bhubaneswar for carrying out our computational work. We also acknowledge Sourin Das, Alexander A. Zyuzin and S. D. Mahanti for stimulating discussions. D.C acknowledges financial support from DST (project number SR/WOS-A/PM-52/2019).


\end{document}